\documentclass[a4paper,fleqn,usenatbib]{mnras}
\usepackage{amsmath}
\usepackage{mathptmx}
\usepackage{txfonts}
\usepackage[T1]{fontenc}
\usepackage{graphicx}	% Including figure files
\usepackage{amssymb}	% Extra maths symbols
\usepackage[normalem]{ulem}
\usepackage{longtable}
\usepackage{multirow}

\newcommand{\be}{\begin{equation}}
\newcommand{\e}{\end{equation}}
\newcommand{\bear}{\begin{eqnarray}}
\newcommand{\ear}{\end{eqnarray}}

\newcommand{\hmpc}{{\, h^{-1}\, {\rm Mpc}}}

\def\apjl{ApJL}
\def\apjs{ApJS}

\def\mnras{MNRAS}
\def\aap{A\&A}

\def\apjs{ApJS}
\def\apjl{ApJ Letters}

\title[Environmental dependence of galactic bars]{Do galactic bars
  depend on environment?: An information theoretic analysis of Galaxy
  Zoo 2}

\author[Sarkar, S., Pandey, B., Bhattacharjee, S.] {Suman
  Sarkar$^{1}$\thanks{suman2reach@gmail.com}, Biswajit
  Pandey$^{1}$\thanks{biswap@visva-bharati.ac.in}, Snehasish
  Bhattacharjee$^{2}$\thanks{snehasish.bhattacharjee.666@gmail.com}
  \\$^1$ Department of Physics, Visva-Bharati University,
  Santiniketan, Birbhum, 731235, India \\$^2$Department of Astronomy,
  Osmania University, Hyderabad, 500007, India}

 \date{\today}

 \pubyear{2020}
  
\begin{document}
\label{firstpage}
\pagerange{\pageref{firstpage}--\pageref{lastpage}}      
\maketitle
       
\begin{abstract}
  We use an information theoretic framework to analyze data from the
  Galaxy Zoo 2 project and study if there are any statistically
  significant correlations between the presence of bars in spiral
  galaxies and their environment. We measure the mutual information
  between the barredness of galaxies and their environments in a
  volume limited sample ($M_r \leq -21$) and compare it with the same
  in datasets where (i) the bar/unbar classifications are randomized
  and (ii) the spatial distribution of galaxies are shuffled on
  different length scales.  We assess the statistical significance of
  the differences in the mutual information using a t-test and find
  that both randomization of morphological classifications and
  shuffling of spatial distribution do not alter the mutual
  information in a statistically significant way. The non-zero mutual
  information between barredness and environment arises due to the
  finite and discrete nature of the dataset which can be entirely
  explained by mock Poisson distributions. We also separately compare
  the cumulative distribution functions of the barred and unbarred
  galaxies as a function of their local density.  Using a
  Kolmogorov-Smirnov test, we find that the null hypothesis can not be
  rejected even at $75\%$ confidence level. Our analysis indicates
  that environments do not play a significant role in the formation of
  a bar, which is largely determined by the internal processes of the
  host galaxy.
\end{abstract}


\section{Introduction}
Observations suggest that a significant fraction of spiral galaxies in
the present universe are barred \citep{eskridge, marinovajogee,
  barazza08}. Even our Milky Way is known to host a bar-like structure
\citep{binney, wegg}. Bars are extended linear structures that results
from disc instabilities \citep{toomre64}. The stellar bars transfer
angular momentum to the outer disc \citep{lyndenbell, athanassoula02,
  berentzen07} and also help to redistribute angular momentum between
the disk and the surrounding dark matter halo \citep{weinberg85,
  debattista, athanassoula03, berentzen06}. They are efficient in
driving gas into the central regions of galaxies which can trigger
starbursts and AGN activity and also help to grow a central bulge
component \citep{schwarz81,kormendy82,shlosman89,hunt99, knapen95,
  knapen00, kormendy04,jogee05, laurikanen04, sheth05,
  laurikanen07}. The bars can thus play a driving role in the
evolution of disk galaxies.

It is still not clear if the formation and evolution of bars are
purely governed by internal secular processes. Galaxies form and
evolve in diverse environments in the cosmic web. They form at the
centre of the dark matter halos \citep{white78} which are embedded in
different environments of the cosmic web. The mass, shape and angular
momentum of the dark matter halos are known to be influenced by their
large-scale environments \citep{hahn1, hahn2}. So the environment may
impart an indirect influence on the formation and evolution of
galactic bars.  The different assembly history of the dark matter
halos causes the early-forming low mass halos to cluster more strongly
as compared to the late-forming halos of similar mass \citep{croton,
  gao07}. Such clustering bias of dark matter halos may indirectly
affect the formation of galactic bars. The direct external influence
of environments may also play a role in the formation and evolution of
bars in spiral galaxies. Studies with N-body simulations suggest that
tidal interactions and the passage of a companion galaxy can trigger
the formation of bars in disk galaxies \citep{byrd90, gerin90,
  berentzen04, valpuesta, lokas, ghosh20}. These trends have been also
supported by observations \citep{elmegreen90, giuricin93, mendez12}.

It would be interesting to know whether the formation and evolution of
galactic bars are influenced by their environment. A significant
number of studies have been carried out to test the correlation
between the environment and the presence of bars in spiral
galaxies. \citet{thomson81} studies the radial distribution of barred
galaxies in the Coma cluster and find that a significantly larger
fraction of barred galaxies is found at the cluster
core. \citet{giuricin93} use NGC catalogue to study the effect of
local galaxy density on the presence of bars and find that the early
type and low luminosity spiral galaxies in high density environments
tend to be barred.  \citet{eskridge} find a slightly higher fraction
of barred galaxies in the Fornax and Virgo clusters compared to the
fields. \citet{barway} study bar fraction in lenticular galaxies using
SDSS DR7 and find a higher fraction of barred galaxies in clusters
than in the fields. \citet{skibba12} study the environmental
dependence of bars in spiral galaxies using the Galaxy Zoo 2 project
and find that the redder galaxies with higher stellar mass are more
likely to have bars. They reported a significant bar-environment
correlation which shows that the barred galaxies more frequently occur
in denser environments than their unbarred counterparts.

Contrary to these findings, several studies reported no dependence of
bars on the environment. \citet{vandenbergh} investigate the
dependence of bar frequency in fields, groups and cluster environments
and find no evidence for any role of environment on the formation of
bars. \citet{li09} studied the projected redshift-space two-point
cross-correlation functions of barred and unbarred galaxies in the
SDSS and find that at a fixed stellar mass, the clustering of barred
and unbarred galaxies are indistinguishable over the scales 20 kpc -
30 Mpc. \citet{barazza09} find that the fraction and properties of
bars in clusters and fields are quite similar. \citet{aguerri09} study
bar fraction as a function of local galaxy density using SDSS DR5 and
find that there is no difference in the local galaxy density of barred
and unbarred galaxies. \citet{cameron10} investigate the evolution of
bar fraction in the COSMOS field and reported that the evolution of
the barred galaxy populations does not depend on the large-scale
environmental density. \citet{mendez10} use HST ACS data to study the
bar fraction in the Coma cluster and find that the bar fraction does
not vary significantly in the centre and outskirts of the
cluster. \citet{martinez11} study the relationship between the
fraction of barred spirals and a number of environmental parameters
and find that the fraction of barred spirals is insensitive to their
environment. \citet{lee12} use SDSS DR7 to study the dependence of
bars on environment and find that the fraction of barred galaxies are
independent of their large-scale environment when the other galaxy
properties are fixed. \citet{marinova12} study bars in massive disk
galaxies using data from the HST ACS Treasury survey of the Coma
cluster and find that the bar fraction does not show a statistically
significant variation across environments.

Clearly, the correlation between the occurrence of bars in spirals and
their environment still remains a debated issue and there are no clear
consensus on the bar-environment correlation.

Most of the studies in this field are plagued by smaller size of data
samples which made it difficult to derive statistically meaningful
conclusions. The SDSS \citep{york00} is the largest and most
successful redshift survey to date which has provided the most
detailed three dimensional map of the nearby universe and a wealth of
information about the individual galaxies. The Galaxy Zoo
\citep{lintott1,lintott2} is a citizen science project based on the
SDSS and HST data which invites volunteers to help in the
morphological classification of a large number of galaxies by visual
inspection of their images. We plan to use data from the Galaxy Zoo 2
project \citep{willet13} for the present work.
 
\citet{pandey17} proposed an information theoretic framework to study
the correlation between the morphology of a galaxy and its large-scale
environment. They considered spirals and ellipticals as two distinct
morphological classes and find a synergic interaction between
morphology and environment up to a length scale of $30
\hmpc$. Recently \citet{sarkar20} show that the observed excess mutual
information between morphology and environment are statistically
significant at $99.9\%$ confidence level. Another study by
\citet{bhattacharjee} show that a conditioning on stellar mass does
not explain the statistically significant mutual information between
morphology and environment on larger length scales. Galactic bar is an
important morphological feature based on which a spiral galaxy is
further classified as barred or unbarred. It would be natural to ask
if there exists a bar-environment correlation similar to the
correlation observed between morphology and environment.

We would like to measure the mutual information between barredness of
a galaxy and its environment on different length scales. One can
randomize the information about barredness and also shuffle the
spatial distribution after dividing it into smaller
sub-cubes. Comparing the mutual information in these distributions with
that from the original distribution would allow us to test the
statistical significance of any observed correlation between
barredness and environment \citep{sarkar20}. In this work, we use this
information theoretic framework to test the large-scale environmental
dependence of galactic bars if any.

The local density of galaxies governs the various external influences
such as tidal interactions which may act as an external trigger for
bar formation. Comparing the number density of galaxies at the
locations of the barred and unbarred spirals can elucidate this
issue. Keeping this in mind, we also separately test any effects of
local density on the presence of galactic bars in spiral galaxies.

\section{DATA}
\subsection{SDSS and Galaxy Zoo 2 data}
\begin{figure*}
\resizebox{8 cm}{!}{\rotatebox{0}{\includegraphics{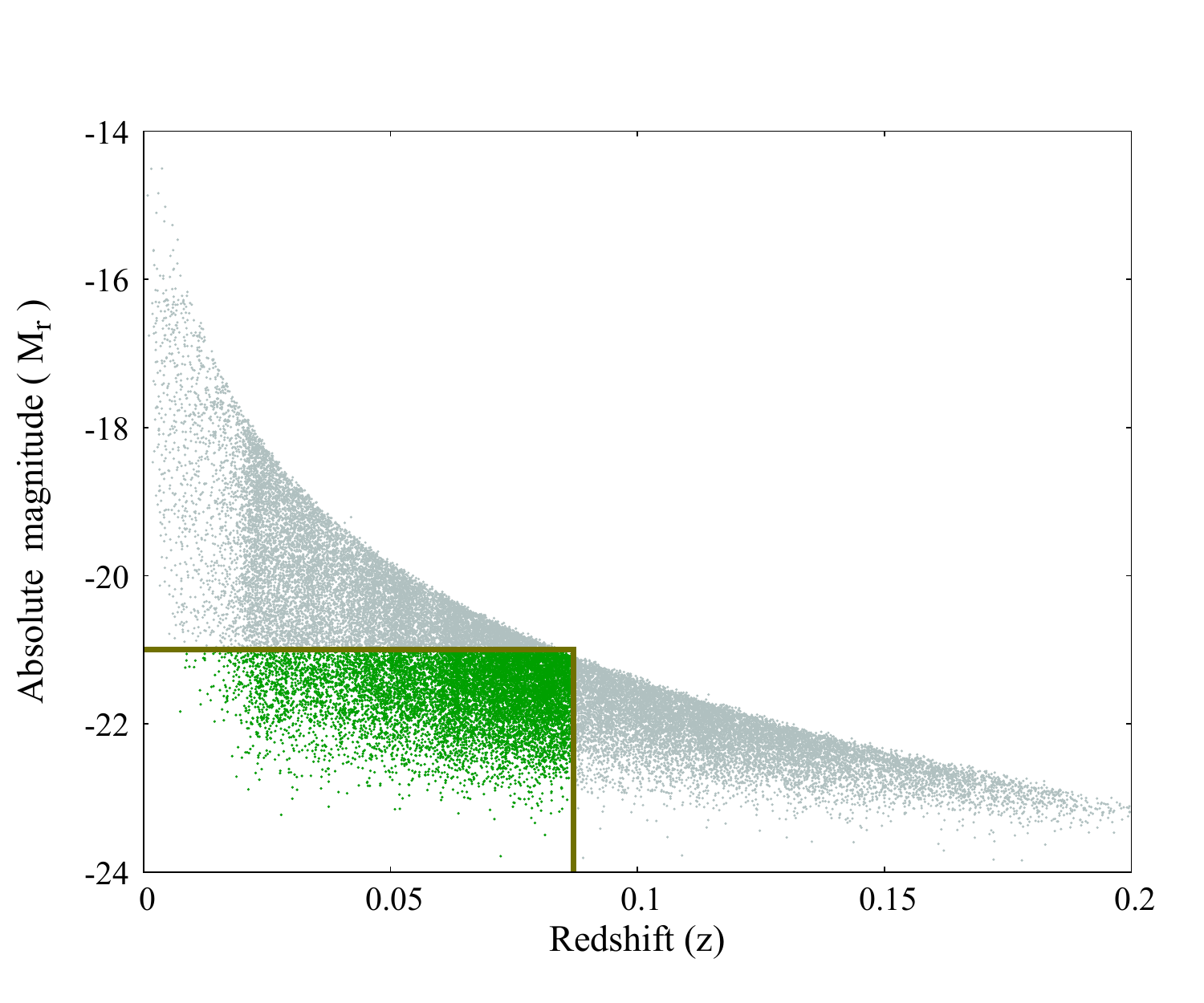}}} \hspace{1 cm}
\resizebox{8 cm}{!}{\rotatebox{0}{\includegraphics{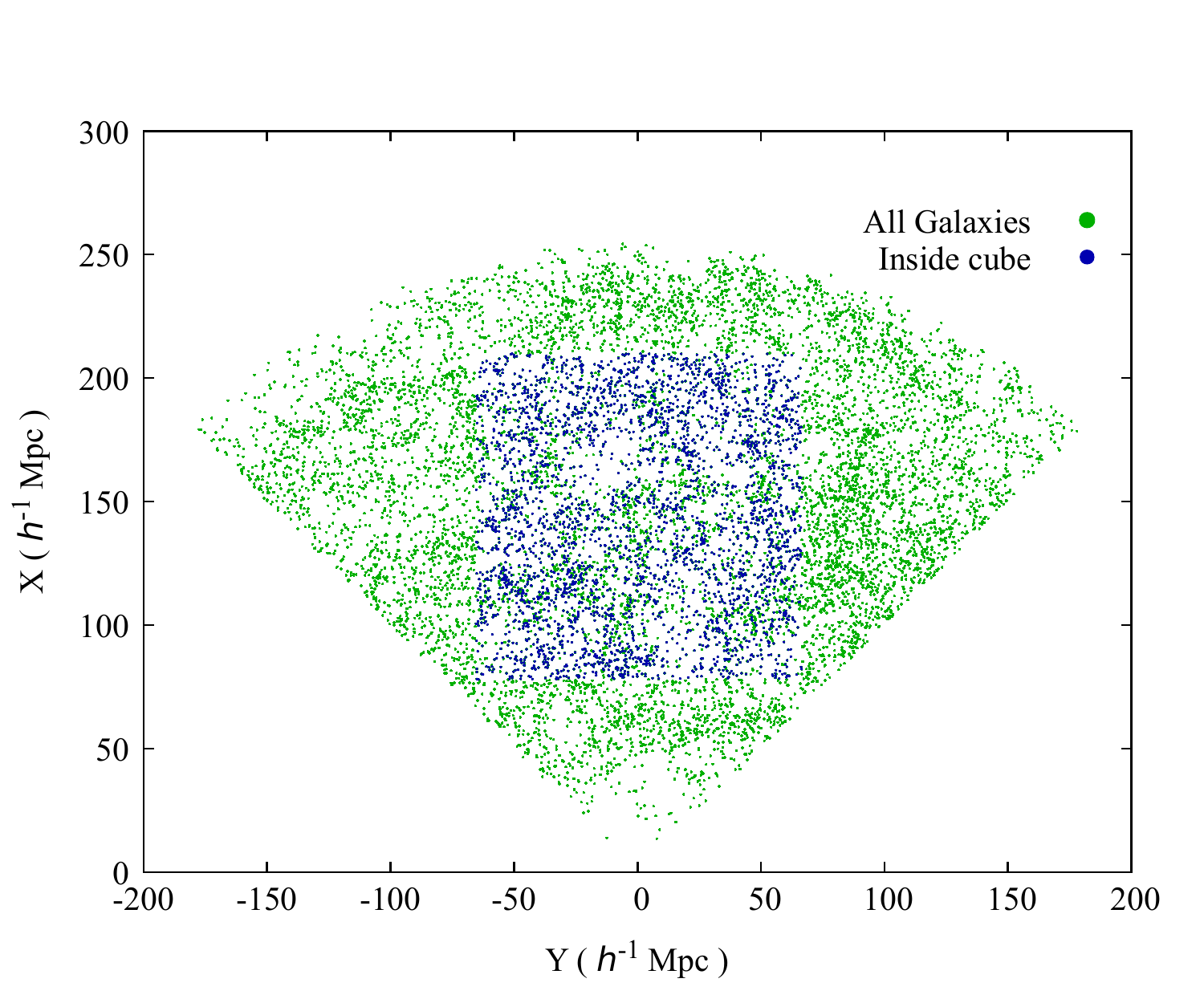}}} \\ 
\resizebox{8 cm}{!}{\rotatebox{0}{\includegraphics{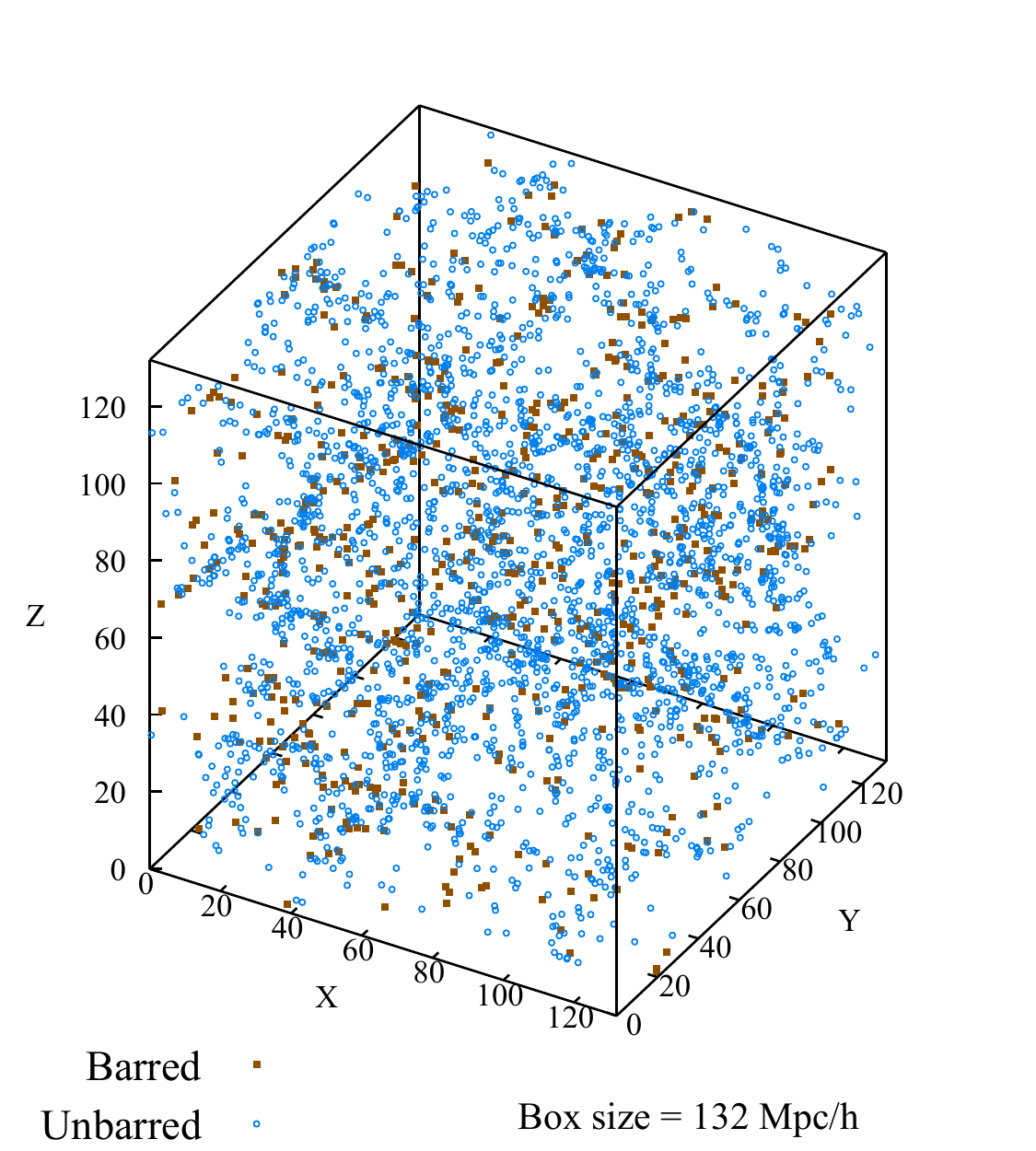}}} \hspace{1 cm}
\resizebox{8 cm}{!}{\rotatebox{0}{\includegraphics{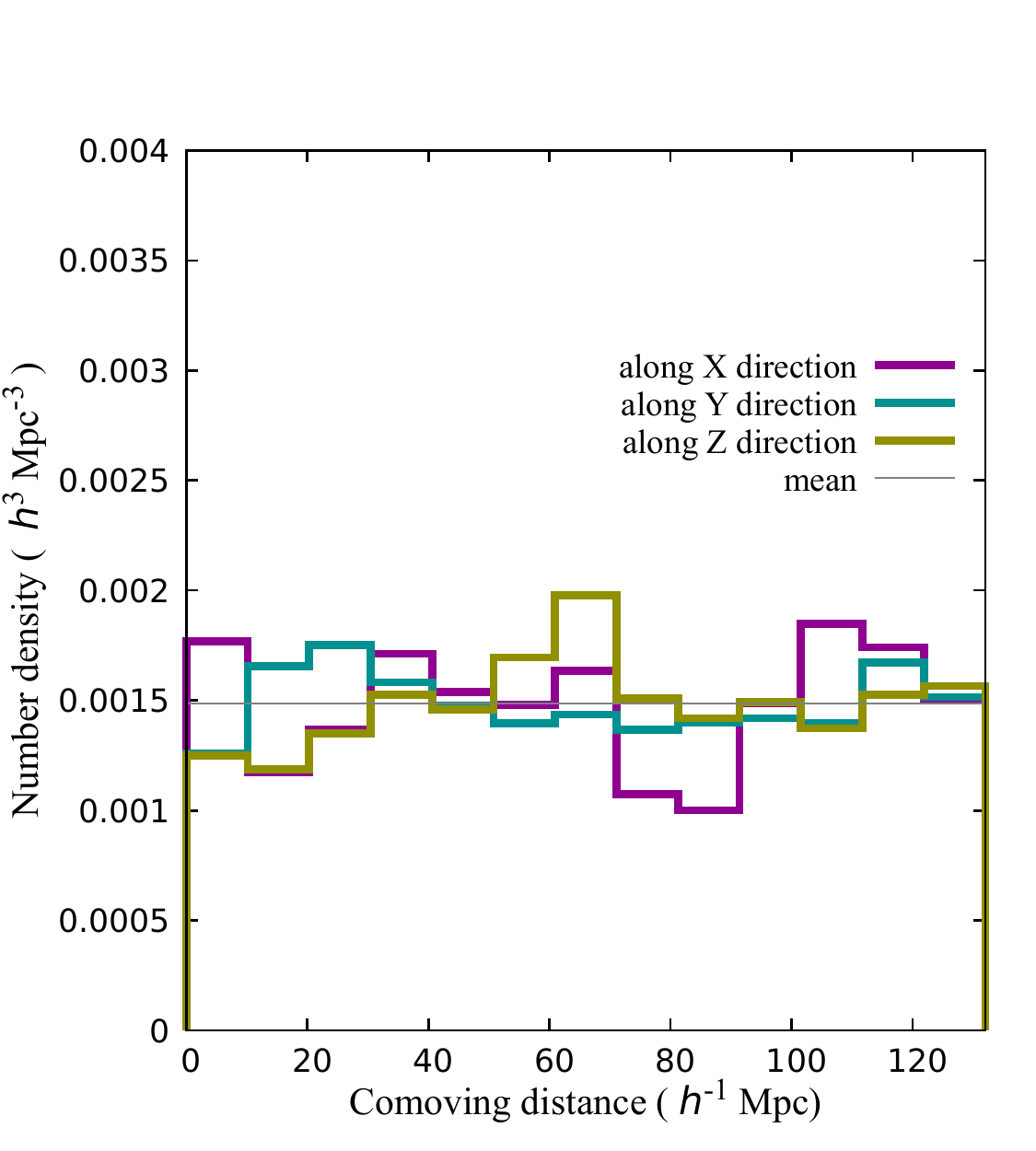}}} \\
\caption{The top left panel shows the definition of the volume limited
  sample used in the present analysis. The projected distribution of
  the galaxies in the volume limited sample is shown in the top right
  panel. The galaxies within the cubic region are shown with a
  different colour in the same panel. The bottom left panel shows the
  spatial distribution of the barred and unbarred spirals in the cubic
  region defined within the volume limited sample. The bottom right
  panel shows the comoving number density of galaxies as a function of
  distance along three different axes within the cubic region. The
  comoving number densities are obtained in uniform slices of
  thickness $10.15 \hmpc$.}
\label{fig:sample}
\end{figure*}

We use data from the Sloan Digital Sky survey (SDSS) for the present
analysis. We use Structured Query Language (SQL) to extract the
required data from the SDSS
SkyServer\footnote{https://skyserver.sdss.org/casjobs/}. The SDSS
\citep{york00} covers $9,376$ square degrees of sky for spectroscopy
where $2,863,635$ galaxies were chosen as targets. The morphological
information of galaxies in the SDSS main-sample \citep{abazajian09} is
provided by Galaxy Zoo 2 (GZ2) \citep{willet13}. GZ2 is the second
phase of the original Galaxy zoo project(GZ1)
\citep{lintott1} \footnote{http://zoo1.galaxyzoo.org} which is a
citizen scientist programme for morphological classification of
galaxies through visual inspection of images. GZ1 provides
morphological classifications of $\sim 900000$ galaxies drawn from the
SDSS. GZ2 targets a subset of $\sim 300000$ galaxies from GZ1 for a
more detailed morphological classifications. The GZ2 decision tree
consists of a total of 11 tasks. It differentiates the galaxies having
a disk from the smooth (E/S0) ones and also record the various
prominent features of the galaxies like presence of bars, number of
spiral arms, arc or lens shapes. We combine the $zoo2Mainspecz$ table
with $specobjAll$ and $photoz$ to retrieve the required
information. We use a critical value of debiased vote fraction to
select the spirals ($ t04\_spiral\_a08\_spiral\_debiased > 0.6$). The
barred and unbarred spirals are selected using a similar cut-off in
the value of the debiased vote fraction
($t03\_bar\_a06\_bar\_debiased>0.6$ and
$t03\_bar\_a07\_no\_bar\_debiased >0.6$). The cut-off values for the
debiased vote fractions were chosen so as to have a reasonable number
of galaxies in the volume limited sample to be prepared.  We identify
a contiguous region in the northern galactic hemisphere and select all
the classified barred and unbarred spirals between right ascension
$135^\circ$ and $255^\circ$ and declination $0^\circ$ and
$60^\circ$. We prepare a volume limited sample by restricting the {\it
  r}-band Petrosian absolute magnitude to $M_r \leq -21$. The galactic
extinction corrected {\it r}-band Petrosian apparent magnitude limit
of the sample is $m_r <17$. We get a volume limited sample which
extends upto redshift $z \leq 0.087$ and contains a total $11260$
galaxies ($2214$ barred and $9046$ unbarred). We then extract all the
galaxies within a cubic region of sides $132 \hmpc$ from the volume
limited sample. This is the largest cube that can fit within the
volume limited sample. The resulting cube contains a total $3420$
galaxies of which $690$ are barred and $2730$ are
unbarred. We show the definition of the volume limited
  sample, the spatial distributions of the galaxies in the sample and
  the variations of comoving number density in it in
  \autoref{fig:sample}.\\

We use the $\Lambda$CDM cosmological model with $\Omega_{m0}=0.315$,
$\Omega_{\Lambda0}=0.685$ and $h=0.674$ \citep{planck18} for
conversion of redshifts to comoving distances.
\subsection{Mock Poisson samples}
We generate $10$ mock Poisson samples each with $3420$ points
distributed within a cubic region of side $132 \hmpc$. We randomly
label $690$ points as barred and $2730$ points as unbarred in each of
these distributions. These mock data sets have the same number of
galaxies as in the actual SDSS data cube. The ratio of barred to
unbarred spirals in the mock samples are also kept same as the actual
data.

\section{Method of analysis}
We now have a magnitude limited sample of the spiral galaxies defined
within a cubic region of size $L\hmpc$. We divide this cube in
$N_d=n_g^3$ voxels with size $d=(\frac{L}{n_g})\hmpc$. One can have
different sets of voxels by changing the number of grids
$n_g$. Different choices of $n_g$ allow us to study the environment on
different length scales.\\
 
Let us now define a discrete random variable $X$ that represents the
environment at a certain length scale. The probability of finding a
randomly selected galaxy in the $i^{th}$ voxel will be
$p(X_i)=\frac{N_i}{N}$, where $N_i$ is the number of galaxies in the
$i^{th}$ voxel and $N=\sum_{i=1}^{N_d}{N_i}$ is the total number of
galaxies in the entire cube. The random variable $X$ have $N_d$
outcomes ($\{X_i: i=1,2,....N_d\}$).

The information entropy \citep{shannon48} associated with the random
variable $X$ on length scale of $d \hmpc$ is given by,

\begin{eqnarray}
H(X)& = &-\sum_{i=1}^{N_d} p(X_i) \log p(X_i) \nonumber \\
&=&\log N - \sum_{i=1}^{N_d} \frac{N_i \log N_i}{N}.
  \label{eqn:Hx}
\end{eqnarray}

Let us now define another variable $Y$ that represents the presence or
absence of bar in spiral galaxies. Here, $Y$ only takes two values
$Y_1$ for barred \& $Y_2$ for unbarred. If there are $N_1$ barred and
$N_2$ unbarred galaxies in the cube then the Shannon entropy
associated with the variable $Y$ will be

\begin{eqnarray}
H(Y)& = &- \left( \frac{N_{1}}{N} \log \frac{N_{1}}{N} +
\frac{N_{2}}{N} \log \frac{N_{2}}{N} \right) \nonumber \\ &=& \log N-
\frac{ N_{1} \log N_{1} + N_{2} \log N_{2}}{N}
  \label{eqn:Hy}
\end{eqnarray}
 
Here, $N=N_1+N_2$ is the total number of galaxies in the sample. Any
variation in the voxel size will not change the number of barred and
unbarred galaxies and the value of $H(Y)$ is independent of the grid
size.

We also calculate the joint entropy for the variables $X$ and
$Y$. Which is given by,
 
\begin{eqnarray}
H(X,Y) &=& -\sum_{i=1}^{N_d} \sum_{j=2}^{2} p(X_i,Y_j) \log p(X_i,Y_j) \nonumber \\
&=&\log N - \frac{1}{N}\sum_{i=1}^{N_d} \sum_{j=1}^{2} N_{ij} \log N_{ij}. 
\label{eqn:Hxy}
\end{eqnarray}

Here $N_{ij}$ is the number of galaxies that resides in the $i^{th}$
voxel and belongs to the $j^{th}$ morphology. So we have,
\begin{eqnarray}
\sum_{i=1}^{N_d}{\sum_{j=1}^{2}{N_{ij}}}=N \nonumber .
\label{eq:nij}
\end{eqnarray}

If the two variables $X$ and $Y$ are uncorrelated then
$H(X)+H(Y)=H(X,Y)$. Otherwise the joint entropy would be smaller than
the some of the individual entropies, i.e.  $H(X,Y)<H(X)+H(Y)$.\\

We calculate the mutual information between the two variables $X$ and
$Y$ as,
\begin{eqnarray}
I(X;Y) & = & H(X)+H(Y)-H(X,Y)
  \label{eqn:Ixy}
\end{eqnarray}

The mutual information measures the amount of information shared
between two random variables. In other words, it is the reduction in
uncertainty in the outcome of one random variable due to the
pre-existing knowledge of the other. Higher the mutual information,
greater is the association between the two variables. The mutual
information measures the association between two random variables
irrespective of the nature of the random variables and their
relationship.

\subsection{Randomizing the classification of barred and unbarred galaxies}
We temporarily obliterate the actual bar/unbar classifications of the
spiral galaxies in the cube and randomly tag each of them as barred or
unbarred. We do this in such a way that the total number of barred and
unbarred galaxies in the resulting distribution remains the same as
before. Such randomization of classifications would not change $H(X)$
or $H(Y)$ but the joint entropy $H(X,Y)$ of the resulting distribution
is expected to change when the variables are correlated. The
randomization procedure destroys any existing correlations between $X$
and $Y$ turning them into independent random variables. So any mutual
information of physical origin should ideally diminish to zero after
the randomization. We generate 10 randomized datasets from the actual
data for our analysis.

\subsection{Shuffling the spatial distribution}
The SDSS data cube of sides $L = 132\hmpc$ is divided into
$N_c={n_s}^3$ sub cubes of size $l_s=\frac{L}{n_s}$, where $n_s$ is
the number of segments on each side. We shuffle \citep{bhavsar88} the
sub-cubes to obtain a new distribution which contains same number of
galaxies distributed within the same volume. This will destroy any
existing correlations between the environment and the barredness
beyond the size of the sub-cubes used for shuffling. A detail
description of the shuffling procedure can be found in
\cite{sarkar20}. The sub cubes are randomly interchanged with random
rotations in multiples of $90^{\circ}$ along any of the three axes. We
repeated this process for $100 \times N_c$ times so as to allow each
of the sub-cubes to swap its position multiple times, with other
sub-cubes. The shuffling exercise is performed for three different
sizes of shuffling unit, $n_s=3$, $n_s=7$ and $n_s=11$ which
corresponds to $l_s = 44 \hmpc$, $l_s \sim 19 \hmpc$ and $l_s = 12
\hmpc$ respectively. In order to avoid any spurious correlations, we
chose the size of the shuffling units to be different from the grid
sizes used for the estimation of mutual information. We also ensure
that the size of the shuffling units are not integral multiple of grid
sizes and vice versa. We generate 10 shuffled realizations from the
original SDSS data for the present analysis.

\subsection{Testing statistical significance of mutual information with {\it t}-test}
An equal variance $t$-test is used to estimate the statistical
significance of the mutual information between environment and
barredness of the spiral galaxies in the actual SDSS data. At each
length scale, we compare the mutual information obtained for
randomized or shuffled data with that from the original SDSS data. The
$t$-score at each length scale is given by,
\begin{eqnarray}
t= \frac{|{\mu_1}-{\mu_2}|}{\sigma_s \sqrt{\frac{1}{n_1}+\frac{1}{n_2}}}.
\label{eqn:ttest}
\end{eqnarray}

Here ${\mu_1}$ and ${\mu_2}$ are the mean values and $\sigma_1$ and
$\sigma_2$ are the standard deviations at a given length scale for the
two data sets under consideration. $\sigma_s =
\sqrt{\frac{(n_1-1)\sigma_1^2+(n_2-1)\sigma_2^2}{n_1+n_2-2}}$, where
$n_1$ and $n_2$ are the number of samples used to estimate mean and
standard deviation of the two data sets. $(n_1+n_2-2)$ is the degree
of freedom associated with the test.

\begin{figure*}
\resizebox{8 cm}{!}{\rotatebox{0}{\includegraphics{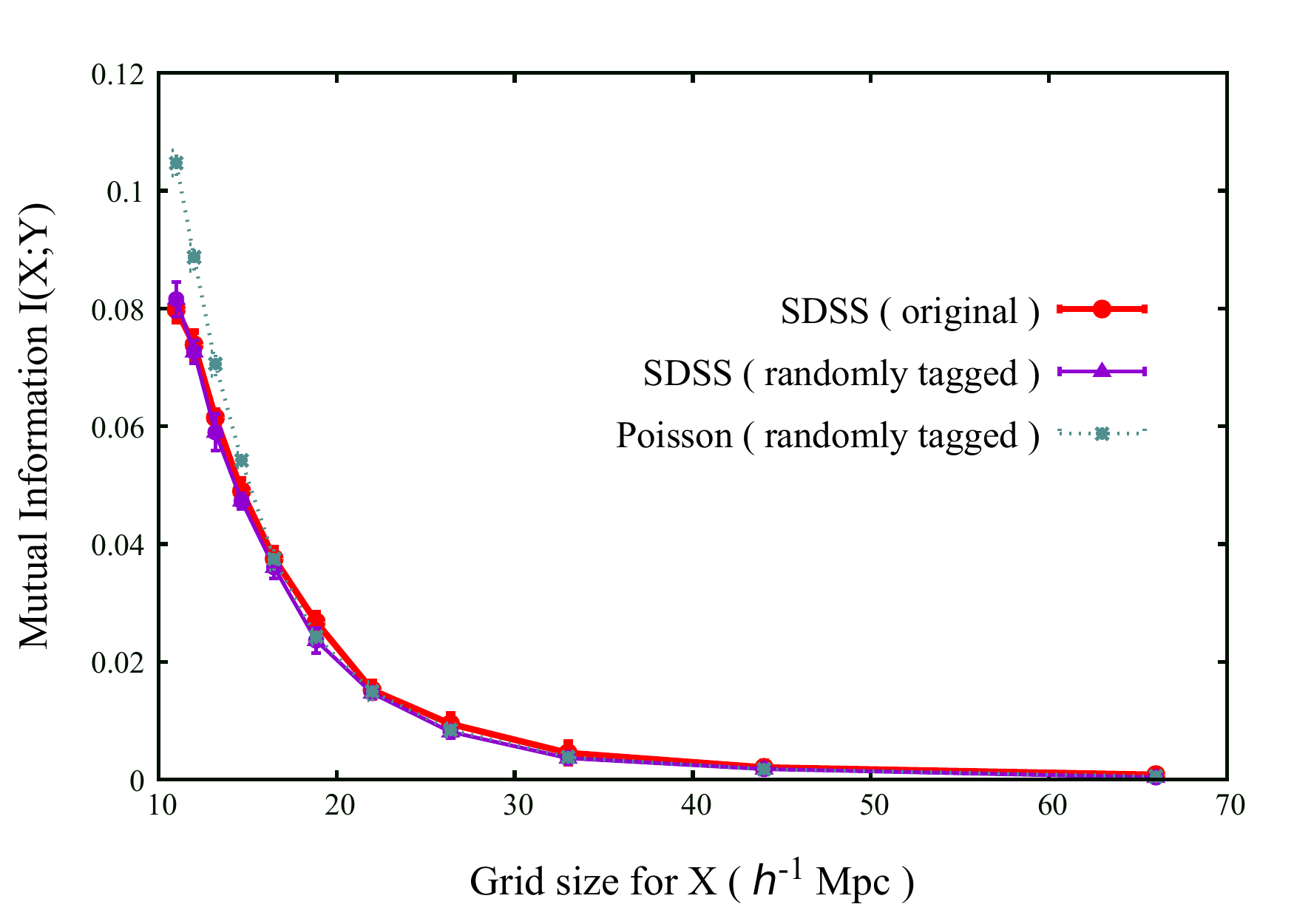}}} \hspace{1 cm}
\resizebox{8 cm}{!}{\rotatebox{0}{\includegraphics{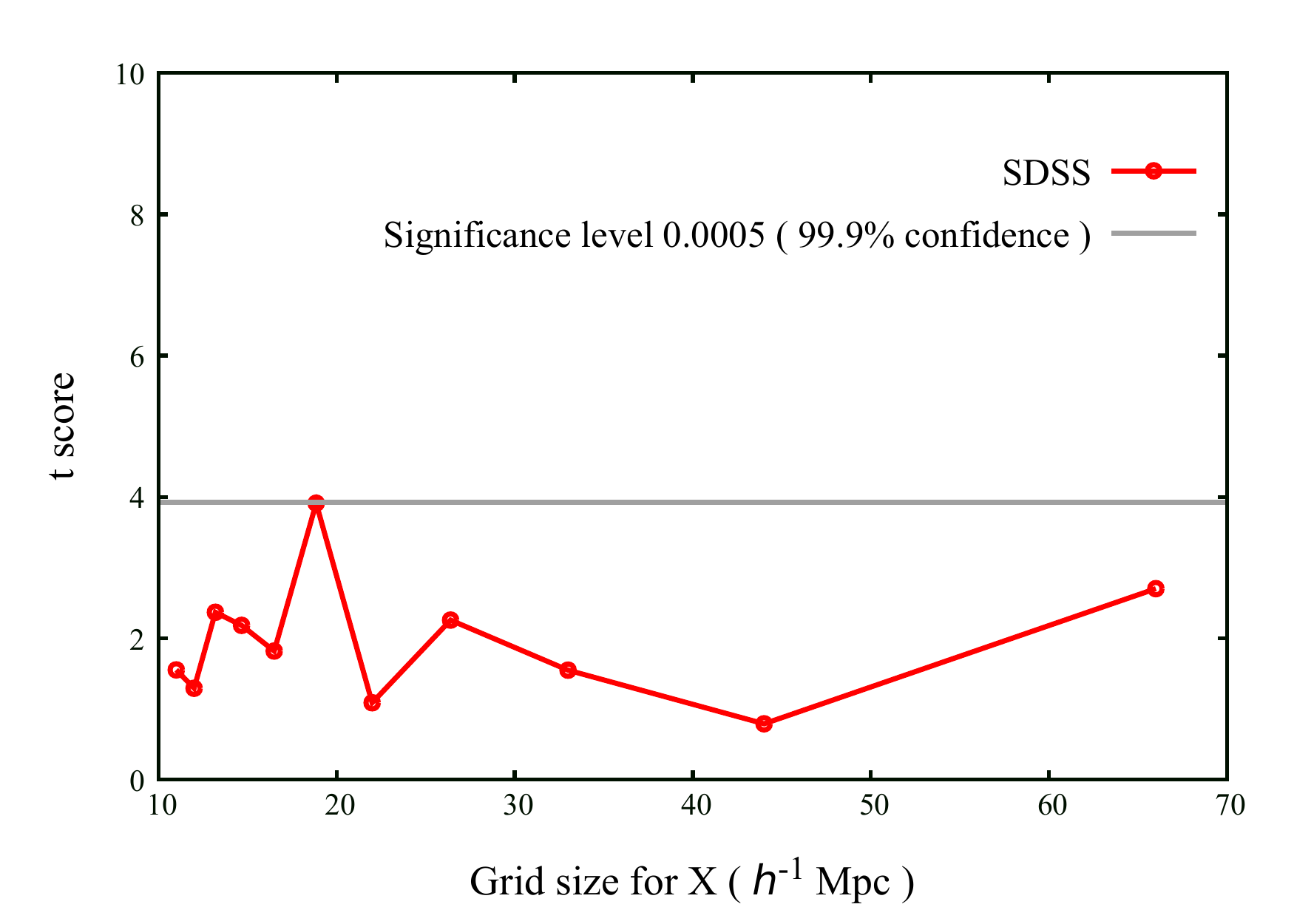}}} \\
\resizebox{8 cm}{!}{\rotatebox{0}{\includegraphics{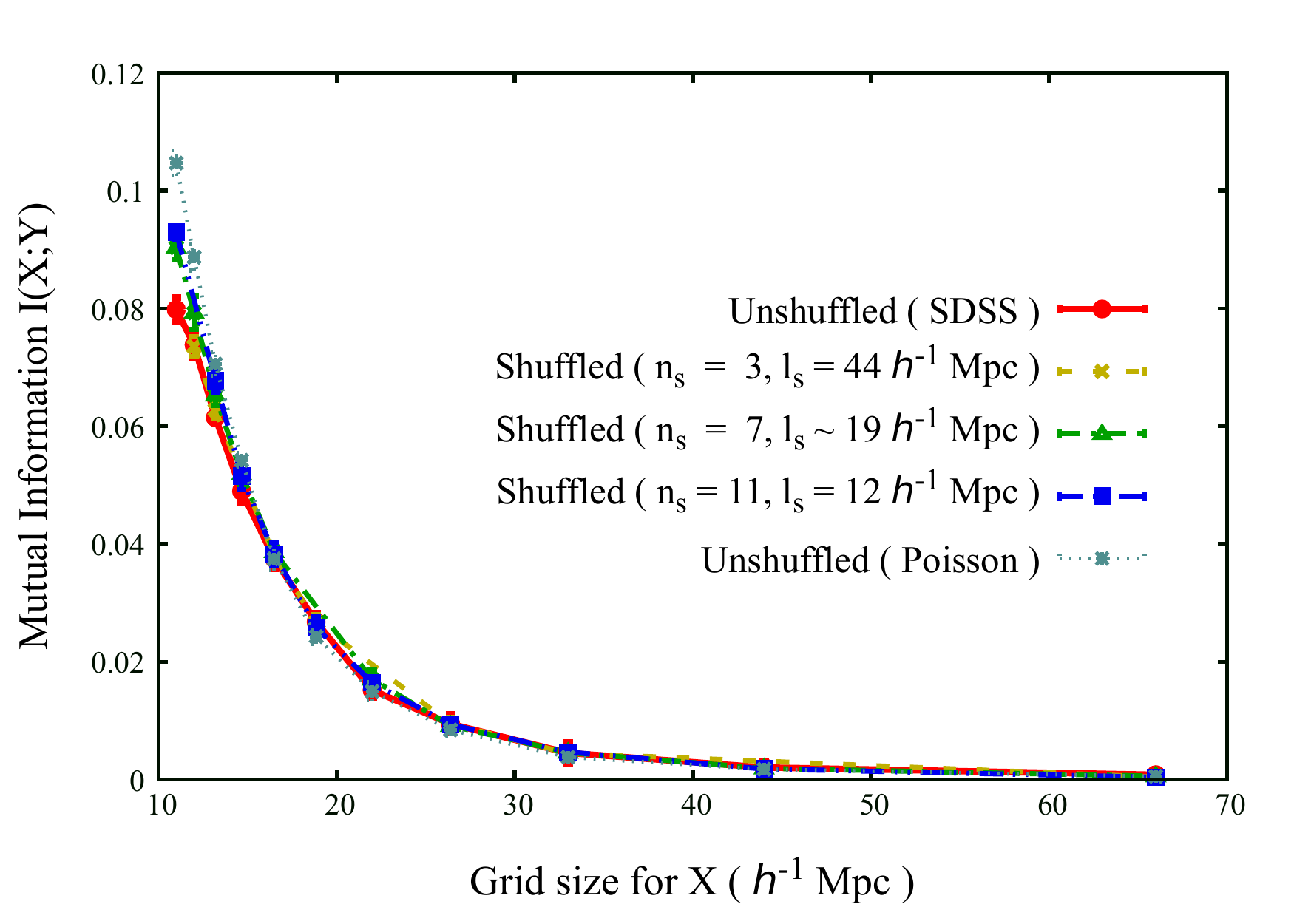}}} \hspace{1 cm}
\resizebox{8 cm}{!}{\rotatebox{0}{\includegraphics{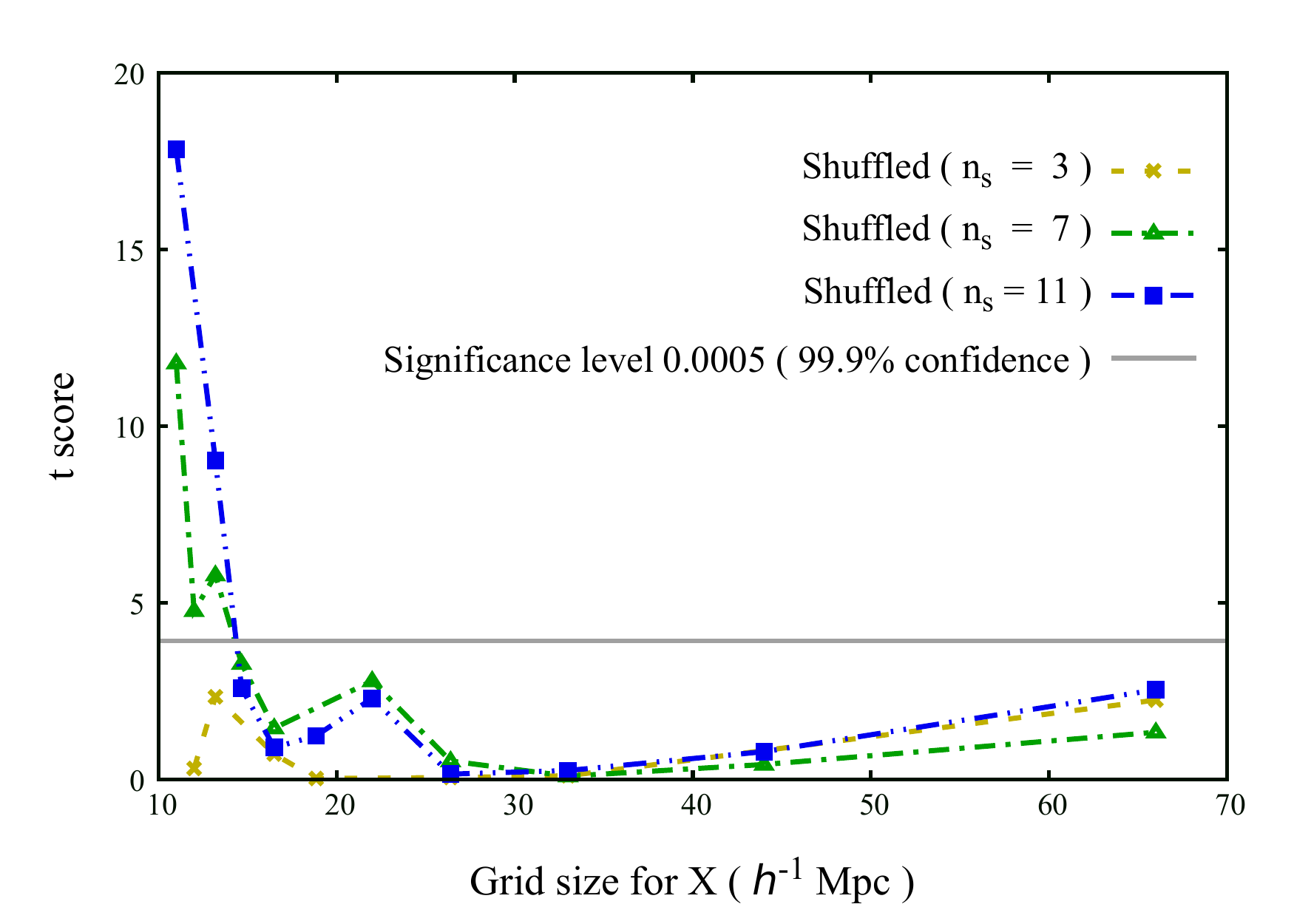}}} \\ 
\caption{The top left panel shows the mutual information between
  barredness and environment as a function of length scales in the
  original SDSS data cube, mock Poisson data cubes and SDSS data cubes
  with randomized bar/unbar classification. The $1-\sigma$ errorbars
  for the SDSS (randomized) and Poisson random datasets are obtained
  using 10 different realizations for each. We estimate the $1-\sigma$
  errorbars for the original SDSS data using $10$ jack-knife samples
  drawn from the original dataset. The top right panel shows the t
  score between original and randomized SDSS data at different length
  scales. The bottom left and bottom right panels of this figure show
  the same but for the shuffled SDSS data along with original SDSS
  data and mock Poisson datasets. The spatial distribution of galaxies
  within the SDSS data cube is shuffled with three different shuffling
  lengths and the corresponding results are shown together in the
  bottom left and bottom right panel of this figure. In each case, we
  ensure that the size of the sub-cubes used for shuffling the data, is
  not equal or integral multiple of the grid sizes used for
  calculating the mutual information.}
\label{fig:ranshuff}
\end{figure*}

\begin{figure*}
\resizebox{10 cm}{!}{\rotatebox{0}{\includegraphics{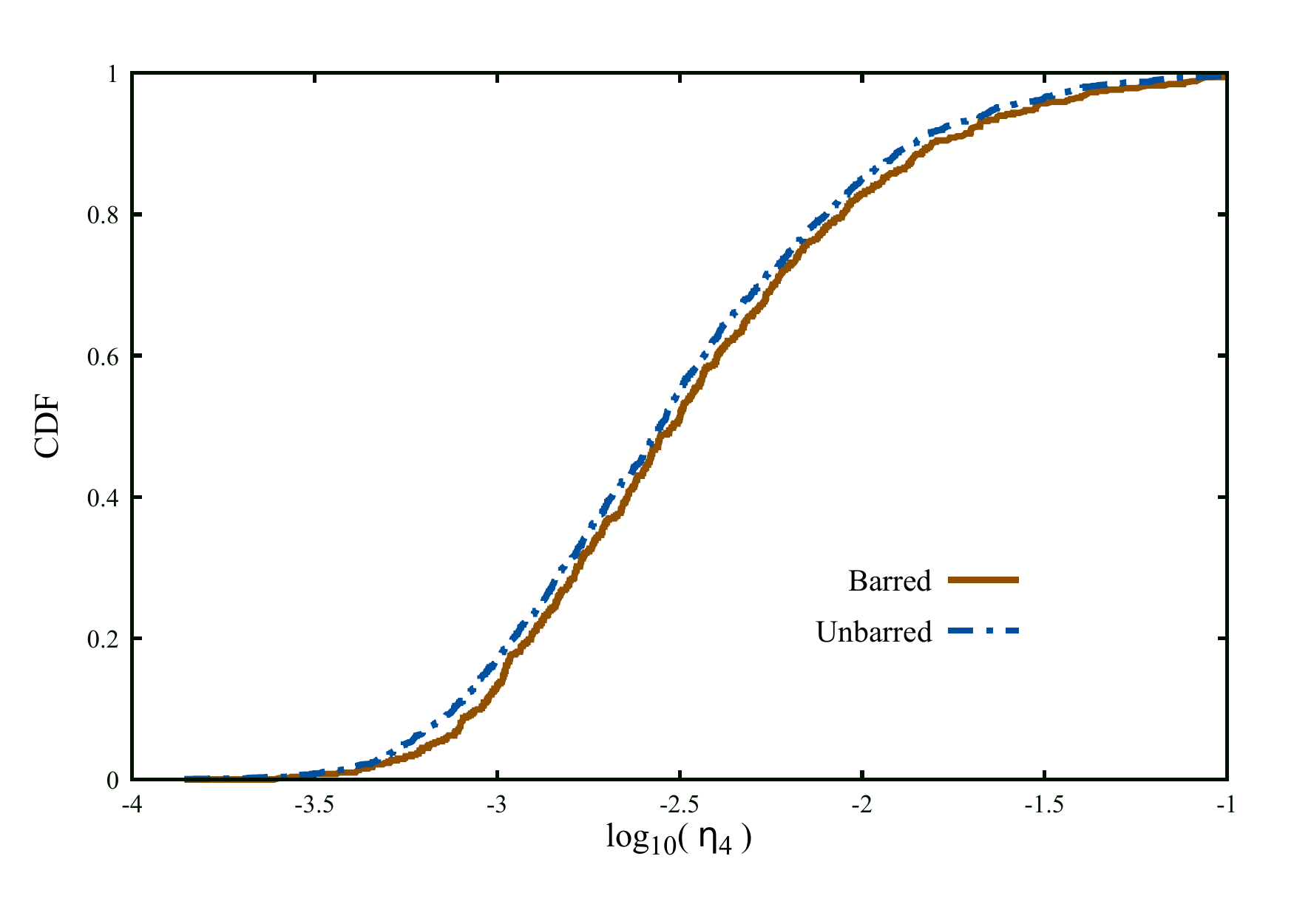}}}
\caption{This shows the cumulative distribution function of barred and
  unbarred spirals as a function of local galaxy density.}
  \label{fig:cdf}
\end{figure*}

\begin{table*}{}
  \caption{This table shows the $t$ scores between actual SDSS data and
    SDSS data with randomized bar/unbar classifications, at different
    length scales. The $p$ values associated with the $t$ scores are
    also listed in the same table.}
\label{tab:trandom}
\begin{tabular}{ccc}
\hline
Grid size ( $\hmpc$ )  & $t$ score & $p$ value \\
\hline
11.00 & 1.551 & $6.91 \times 10^{-2}$ \\
12.00 & 1.292 & $1.06 \times 10^{-1}$ \\
13.20 & 2.366 & $1.47 \times 10^{-2}$ \\
14.67 & 2.182 & $2.13 \times 10^{-2}$ \\
16.50 & 1.819 & $4.28 \times 10^{-2}$ \\
18.86 & 3.912 & $5.11 \times 10^{-4}$ \\
22.00 & 1.086 & $1.46 \times 10^{-1}$ \\
26.40 & 2.255 & $1.84 \times 10^{-2}$ \\
33.00 & 1.547 & $6.96 \times 10^{-2}$ \\
44.00 & 0.788 & $2.20 \times 10^{-1}$ \\
66.00 & 2.700 & $7.32 \times 10^{-3}$ \\
\hline
\end{tabular}
\end{table*}  

\begin{table*}{}
\caption{ This table shows the $t$ scores between original SDSS data
  and shuffled versions of the SDSS data, at different length scales.
  Three different shuffling length scales were used in the analysis.
  We ensure that for each $n_s$, the size of the sub-cubes employed
  for shuffling the data is not equal or integral multiple of the grid
  size used for the estimation of mutual information. The $p$ values
  associated with the $t$ scores are also tabulated here. }
\label{tab:tshuffle}
\begin{tabular}{cccccccc}
\hline
 Grid size   & \multicolumn{2}{c}{$n_s = 3$}  & \multicolumn{2}{c}{$n_s = 7$}  & \multicolumn{2}{c}{$n_s = 11$}\\
( $\hmpc$ ) & $t$ score & $p$ value & $t$ score & $p$ value & $t$ score & $p$ value 	  \\
 
\hline
11.00 &		-	& 		-				& 11.785 	& $3.37 \times 10^{-10}$	& 17.841	& $3.42 \times 10^{-13}$\\
12.00 & 0.314	& $3.79 \times 10^{-1}$	&  4.775 	& $7.56 \times 10^{-5}$ 	&	-		&	- 					\\
13.20 & 2.341	& $1.55 \times 10^{-2}$	&  5.783 	& $8.82 \times 10^{-6}$ 	& 9.027 	& $2.10 \times 10^{-8}$	\\
14.67 & 	-	&		-				&  3.281 	& $2.08 \times 10^{-3}$ 	& 2.590 	& $9.24 \times 10^{-3}$	\\
16.50 & 0.721	& $2.40 \times 10^{-1}$	&  1.452 	& $8.19 \times 10^{-2}$ 	& 0.903 	& $1.89 \times 10^{-1}$	\\
18.86 & 0.029	& $4.89 \times 10^{-1}$	&	-		&			-				& 1.239 	& $1.16 \times 10^{-1}$	\\
22.00 & 	-	& 		-				&  2.792 	& $6.02 \times 10^{-3}$ 	& 2.295 	& $1.70 \times 10^{-2}$	\\
26.40 & 0.056	& $4.78 \times 10^{-1}$	&  0.524 	& $3.03 \times 10^{-1}$		& 0.154 	& $4.40 \times 10^{-1}$	\\
33.00 & 0.104	& $4.59 \times 10^{-1}$	&  0.095 	& $4.63 \times 10^{-1}$		& 0.254 	& $4.01 \times 10^{-1}$	\\
44.00 & 	-	& 		-				&  0.424 	& $3.38 \times 10^{-1}$		& 0.787 	& $2.21 \times 10^{-1}$	\\
66.00 & 2.250	& $1.86 \times 10^{-2}$	&  1.336 	& $9.91 \times 10^{-2}$		& 2.539 	& $1.03 \times 10^{-2}$	\\
\hline
\end{tabular}
\end{table*}

\begin{table*}{}
\caption{This table lists the critical values $D_{KS}(\alpha)$ for
  different significance level $\alpha$ in the Kolmogorov-Smirnov test}
\label{tab:ksp}
\begin{tabular}{lccccc}
\hline
significance level ($\alpha$) & 0.005 & 0.01 & 0.05 & 0.1 & 0.25 \\  
\hline
Confidence level  & 99.5\% & 99\% & 95\% & 90\% & 75\% \\
\hline
$D_{KS}(\alpha)$ & 0.085584 & 0.080481 & 0.067154 & 0.060517 & 0.050420  \\
\hline
\end{tabular}
\end{table*}

\subsection{Local density of barred and unbarred spirals}
We find the distance to the $k^{th}$ nearest neighbour for each
galaxies in the cube and estimate the local number density of galaxies
around it. The $k^{th}$ nearest neighbour density \citep{casertano85}
around a galaxy is defined as
\begin{eqnarray}
\eta_k = \frac{k-1}{V(r_k)}  
\label{eqn:knn}
\end{eqnarray}  
Here $r_k$ is the distance to the $k^{th}$ nearest neighbour and
$V(r_k)=\frac{4}{3}\pi r_k^3$ is the volume of the sphere having a
radius $r_k$. In this work, we have used $k=4$. The value of $r_k$
could be overestimated for the galaxies near the boundary of the cube.
Consequently, the local density of the galaxies near the boundary
could be underestimated. We calculate the local density for only those
galaxies which have $r_k < r_b$, where $r_b$ is the closest distance
of the galaxy from the boundary wall. After using this criteria we are
left with $N_{1}^{'}=514$ barred galaxies and $N_{2}^{'}=2002$
unbarred galaxies. We estimate the densities at the locations of the
barred and unbarred galaxies using \autoref{eqn:knn}. The local
density is estimated in units of $h^3 \, \rm{Mpc}^{-3}$.

\subsection{Testing the difference in the local density of barred and unbarred spirals with Kolmogorov-Smirnov test}
We use the two-sample Kolmogorov-Smirnov test to compare the
cumulative distributions of density for the barred and unbarred
spirals. The Kolmogorov-Smirnov test is a non-parametric test which
makes no assumption about the distributions. The null hypothesis
assumes that both barred and unbarred galaxies are sampled from
populations with identical distributions. We calculate the maximum
difference between the two cumulative distributions. The supremum
difference between the two cumulative distribution functions $D_{KS}$
is defined as
\begin{eqnarray}
D_{KS} & = & \sup_{\eta_{k}} \, \, \{ \,\, | f_{1,
  m}(\eta_{k})-f_{2,m}(\eta_{k}) | \,\, \}
  \label{eqn:Dks}
\end{eqnarray}
$f_{1,m}(\eta_{k})$ and $f_{2,m}(\eta_{k})$ are the cumulative
distribution functions for barred and unbarred spirals at the $m^{th}$
bin where $m \in \{1,2,3...., N^{'} \}$. $\sup$ is the operator that
finds the supremum of all the ($N_{1}^{'}+N_{2}^{'}$)
differences. Here $\sum_{m=1}^{N^{'}}{f_{1,
    m}(\eta_{k})}=\sum_{m=1}^{N^{'}}{f_{2, m}(\eta_{k})}=1$.

The critical value of $D_{KS}$ for a given significance level
($\alpha$) is given by,
 \begin{eqnarray}
D_{KS} (\alpha) & = & \sqrt{- \ln \left( \frac{\alpha}{2} \right) \,\,
  \, \frac{ N_{1}^{'} + N_{2}^{'}}{2 N_{1}^{'} N_{2}^{'}}}
\label{eqn:aks}
\end{eqnarray}
where $N_{1}^{'}$ and $N_{2}^{'}$ are the number of barred and
unbarred spirals in the sample. If $D_{KS}>D_{KS} (\alpha)$ then the
null hypothesis can be rejected at level $\alpha$. We test the null
hypothesis at different significance level to find if the
distributions of barred and unbarred spirals are same or different.

\section{Results}

\subsection{Effects of large-scale environments on galactic bars}
We compare the mutual information in the randomized and shuffled data
sets to that with the original SDSS data in \autoref{fig:ranshuff}.
The mutual information between barredness and environment in the SDSS
data is shown in the top left panel of \autoref{fig:ranshuff}. We find
a non-zero mutual information between the barredness and the
environment in the SDSS data, which decreases with increasing length
scales. We would like to test the statistical significance of these
non-zero mutual information. When we compare the mutual information in
the original data with the randomized data sets, we find that the
randomization of barred/unbarred classifications have no impact on the
mutual information between barredness and environment. We also compare
our results to mock Poisson distributions to understand the relevance
of the observed non-zero mutual information. We find that the observed
mutual information in the SDSS and Poisson data sets are nearly
identical beyond $15 \hmpc$. The mock Poisson samples show a
relatively higher mutual information than the original and randomized
data at length scales below $15 \hmpc$. This may purely arise due to
the dominance of Poisson noise on smaller length scales. This suggests
that the observed non-zero mutual information originate from the
finite and discrete character of the distributions. We use an equal
variance Student's t-test to identify any statistically significant
differences between the original and randomized data sets. The
t-scores at different length scales together with the critical value
of t-score at $99.9 \%$ confidence level are shown in the top right
panel of \autoref{fig:ranshuff}. We tabulate the t-score and the
associated p-value at each length scale in \autoref{tab:trandom}. The
results show that the null hypothesis can not be rejected at this
confidence interval. This analysis suggests that the large-scale
environment of galaxies have no influence on galactic bars.

We then compare the mutual information in the original and shuffled
data sets in the bottom left panel of \autoref{fig:ranshuff}. The plot
shows that shufffling the spatial distribution of galaxies with three
different shuffling lengths have little to no influence on the mutual
information. The process of shuffling the data is expected to destroy
the mutual information at all length scales beyond the shuffling
length. However, we do not observe any such decrease of mutual
information when the data is shuffled on different length scales. We
notice an increase in the mutual information in shuffled data sets at
the smallest length scales. This is contrary to what one would expect
in a shuffled distribution. When we compare our results with that from
mock Poisson distribution, it shows that an even higher mutual
information is observed in the Poisson distributions at the smallest
length scale. Shuffling the data randomizes the spatial distribution
of galaxies and enhances the Poisson character of the distribution. It
may be noted that the mean-intergalactic separation of our sample is
$\sim 9 \hmpc$ and the measurements of mutual information near these
length scales would be dominated by Poisson noise. Evidently, we do
not assign any physical significance to the increase in mutual
information in the shuffled data at smallest length scales. The
t-scores at different length scales in the shuffled data sets are
shown in the bottom right panel of \autoref{fig:ranshuff}. The
critical t-score at $99.9 \%$ confidence level is also shown together
in the same panel. The t-score and the associated p-value at each
length scale for each shuffling length are tabulated in
\autoref{tab:tshuffle}. The results clearly show that the null
hypothesis can not be rejected at $99.9 \%$ confidence
level. Shuffling the spatial distribution on different length scales
do not alter the mutual information between the barredness and the
environment in a statistically significant way. This again suggests
that the large-scale environment do not play a significant role on the
formation of a bar in spiral galaxy.

\subsection{Effects of small-scale environments on galactic bars}
We also test if the local density of barred and unbarred galaxies are
different in a statistically significant way. We estimate the local
densities at the locations of barred and unbarred spiral galaxies. We
compare their cumulative distribution functions as a function of local
density in \autoref{fig:cdf}. We perform a Kolmogorov-Smirnov test to
assess the statistical significance of the differences between the two
distributions. We find the maximum difference between the two
cumulative distribution functions to be $D_{KS}=0.044240$.  The
critical values of $D_{KS}(\alpha)$ for different significance level
$\alpha$ are tabulated in \autoref{tab:ksp}. The null hypothesis can
be rejected if $D_{KS}>D_{KS}(\alpha)$. We find that the null
hypothesis can not be rejected even at $75\%$ confidence level. This
implies that the local density field of the barred and unbarred
galaxies do not differ in a statistically significant manner and they
reside in similar environments on smaller length scales. So the
small-scale environment can not be held accountable for presence or
absence of galactic bars in spiral galaxies.

\section{Conclusions}
Understanding the role of bars in spiral galaxies are central to
understanding their formation. Bars are known to be the most efficient
means to redistribute materials inside a galaxy. It is yet not clearly
known why some spiral galaxies host a bar while others do not. The
detail process of bar formation may be governed by several
factors. Both the internal secular processes and external triggers may
induce the bar formation in a galaxy. Besides, the large-scale
environment and the assembly history of dark matter halos may also
have an influence on the formation of galactic bars. All these
possibilities must be tested against observations to identify the most
influential factors governing the formation of galactic bars.

We have calculated the mutual information between the barredness of a
galaxy and its environment on different length scales using the SDSS
and Galaxy Zoo 2 data. We randomize the bar/unbar classification of
galaxies and measure the mutual information between the barredness and
environment. There are no statistically significant change in the
mutual information between barredness and environment after the
classifications are randomized. We also shuffle the spatial
distribution of SDSS galaxies after dividing it in smaller sub-cubes
and randomly interchanging their spatial locations along with random
rotations. We also do not observe any statistically significant
difference in the mutual information between the barredness and the
environment after the data is shuffled on different length scales.
The analysis do not provide any strong evidence against the null
hypothesis which suggests that the large-scale environment of barred
and unbarred galaxies are similar and there are no correlations
between the barredness of a galaxy and its large-scale environment.

We also separately test any possible influence of local density on the
presence of galactic bars. We measure the local density at the
locations of barred and unbarred galaxies and then compare their
cumulative distribution functions using a Kolmogorov-Smirnov test. The
test favours the null hypothesis which indicates that the local
density of barred and unbarred galaxies are quite similar. A study of
the bar fraction in nearby galaxy clusters suggests that the bar
formation in low-mass galaxies are expected to be more susceptible to
their environment than the bright or massive galaxies
\citep{mendez12}. The volume limited sample analyzed in this work
consists of brighter galaxies for which the bar formation may be
unaffected by their environment.

In the present work, we explore any possible role of the small-scale
and large-scale environments of galaxies on the formation of galactic
bars. Our analysis clearly indicates that the presence or absence of
bars in spiral galaxies do not depend on either their small-scale or
large-scale environments. This suggests that the formation of galactic
bar is largely decided by the internal processes of the host galaxy.

\section{Data availability}
The data underlying this article is available in
https://skyserver.sdss.org/casjobs/ . The datasets were derived from
sources in the public domain: https://www.sdss.org/ and
http://zoo1.galaxyzoo.org .

\section{ACKNOWLEDGEMENT}  
We sincerely thank an anonymous reviewer for insightful comments and
suggestions. The authors would like to thank the SDSS and Galaxy Zoo
team for making the data public. The authors also acknowledge the
efforts of the Galaxy Zoo 2 volunteers for the detailed visual
morphological classifications of the SDSS galaxies, without which this
work would not be possible.

BP would like to acknowledge financial support from the SERB,
DST, Government of India through the project CRG/2019/001110. BP would
also like to acknowledge IUCAA, Pune for providing support through
associateship programme.

Funding for the SDSS and SDSS-II has been provided by the Alfred
P. Sloan Foundation, the Participating Institutions, the National
Science Foundation, the U.S. Department of Energy, the National
Aeronautics and Space Administration, the Japanese Monbukagakusho, the
Max Planck Society, and the Higher Education Funding Council for
England. The SDSS website is http://www.sdss.org/.

The SDSS is managed by the Astrophysical Research Consortium for the
Participating Institutions. The Participating Institutions are the
American Museum of Natural History, Astrophysical Institute Potsdam,
University of Basel, University of Cambridge, Case Western Reserve
University, University of Chicago, Drexel University, Fermilab, the
Institute for Advanced Study, the Japan Participation Group, Johns
Hopkins University, the Joint Institute for Nuclear Astrophysics, the
Kavli Institute for Particle Astrophysics and Cosmology, the Korean
Scientist Group, the Chinese Academy of Sciences (LAMOST), Los Alamos
National Laboratory, the Max-Planck-Institute for Astronomy (MPIA),
the Max-Planck-Institute for Astrophysics (MPA), New Mexico State
University, Ohio State University, University of Pittsburgh,
University of Portsmouth, Princeton University, the United States
Naval Observatory, and the University of Washington.

% Don't change these lines
\bsp	% typesetting comment
\label{lastpage}

\begin{thebibliography}{99}

\bibitem[\protect\citeauthoryear{Abazajian et al.}{2009}]{abazajian09}
  Abazajian K.~N., Adelman-McCarthy J.~K., Ag{\"u}eros M.~A., Allam
  S.~S., Allende Prieto C., An D., Anderson K.~S.~J., et al., 2009,
  ApJS, 182, 543

\bibitem[\protect\citeauthoryear{Aguerri, M{\'e}ndez-Abreu, \&
    Corsini}{2009}]{aguerri09} Aguerri J.~A.~L., M{\'e}ndez-Abreu J.,
  Corsini E.~M., 2009, A\&A, 495, 491

\bibitem[\protect\citeauthoryear{Athanassoula}{2002}]{athanassoula02}
  Athanassoula E., 2002, ApJL, 569, L83

\bibitem[\protect\citeauthoryear{Athanassoula}{2003}]{athanassoula03}
  Athanassoula E., 2003, MNRAS, 341, 1179

\bibitem[\protect\citeauthoryear{Barazza, Jogee, \&
    Marinova}{2008}]{barazza08} Barazza F.~D., Jogee S., Marinova I.,
  2008, ApJ, 675, 1194

\bibitem[\protect\citeauthoryear{Barazza et al.}{2009}]{barazza09}
  Barazza F.~D., Jablonka P., Desai V., Jogee S., Arag{\'o}n-Salamanca
  A., De Lucia G., Saglia R.~P., et al., 2009, A\&A, 497, 713

\bibitem[\protect\citeauthoryear{Barway, Wadadekar, \&
    Kembhavi}{2011}]{barway} Barway S., Wadadekar Y., Kembhavi A.~K.,
  2011, MNRAS, 410, L18

\bibitem[\protect\citeauthoryear{Berentzen et al.}{2004}]{berentzen04}
  Berentzen I., Athanassoula E., Heller C.~H., Fricke K.~J., 2004,
  MNRAS, 347, 220

\bibitem[\protect\citeauthoryear{Berentzen, Shlosman, \&
    Jogee}{2006}]{berentzen06} Berentzen I., Shlosman I., Jogee S.,
  2006, ApJ, 637, 582

\bibitem[\protect\citeauthoryear{Berentzen et al.}{2007}]{berentzen07}
  Berentzen I., Shlosman I., Martinez-Valpuesta I., Heller C.~H.,
  2007, ApJ, 666, 189

\bibitem[\protect\citeauthoryear{Bhattacharjee, Pandey, \&
    Sarkar}{2020}]{bhattacharjee} Bhattacharjee S., Pandey B., Sarkar
  S., 2020, JCAP, 039 (2020)

\bibitem[\protect\citeauthoryear{Bhavsar \& Ling}{1988}]{bhavsar88}
  Bhavsar, S.~P.~\& Ling, E.~N.\ 1988, \apjl, 331, L63

\bibitem[\protect\citeauthoryear{Binney et al.}{1991}]{binney} Binney
  J., Gerhard O.~E., Stark A.~A., Bally J., Uchida K.~I., 1991, MNRAS,
  252, 210

\bibitem[\protect\citeauthoryear{Byrd \& Valtonen}{1990}]{byrd90} Byrd
  G., Valtonen M., 1990, ApJ, 350, 89

\bibitem[\protect\citeauthoryear{Cameron et al.}{2010}]{cameron10}
  Cameron E., Carollo C.~M., Oesch P., Aller M.~C., Bschorr T., Cerulo
  P., Aussel H., et al., 2010, MNRAS, 409, 346

\bibitem[\protect\citeauthoryear{Casertano \& Hut}{1985}]{casertano85}
  Casertano S., Hut P., 1985, ApJ, 298, 80

\bibitem[\protect\citeauthoryear{Croton, Gao \& White}{2007}]{croton}
  Croton D.~J., Gao L., White S.~D.~M., 2007, MNRAS, 374, 1303

\bibitem[\protect\citeauthoryear{Debattista \&
    Sellwood}{2000}]{debattista} Debattista V.~P., Sellwood J.~A.,
  2000, ApJ, 543, 704

\bibitem[\protect\citeauthoryear{Elmegreen, Elmegreen, \&
    Bellin}{1990}]{elmegreen90} Elmegreen D.~M., Elmegreen B.~G.,
  Bellin A.~D., 1990, ApJ, 364, 415

\bibitem[\protect\citeauthoryear{Eskridge et al.}{2000}]{eskridge}
  Eskridge P.~B., Frogel J.~A., Pogge R.~W., Quillen A.~C., Davies
  R.~L., DePoy D.~L., Houdashelt M.~L., et al., 2000, AJ, 119, 536

\bibitem[\protect\citeauthoryear{Gao \& White}{2007}]{gao07} Gao L.,
  White S.~D.~M., 2007, MNRAS, 377, L5

\bibitem[\protect\citeauthoryear{Gerin, Combes, \&
    Athanassoula}{1990}]{gerin90} Gerin M., Combes F., Athanassoula
  E., 1990, A\&A, 230, 37

\bibitem[\protect\citeauthoryear{Ghosh et al.}{2020}]{ghosh20} Ghosh
  S., Saha K., Di Matteo P., Combes F., 2020, arXiv, arXiv:2008.04942

\bibitem[\protect\citeauthoryear{Giuricin et al.}{1993}]{giuricin93}
  Giuricin G., Mardirossian F., Mezzetti M., Monaco P., 1993, ApJ,
  407, 22

\bibitem[Hahn et al.(2007)]{hahn1} Hahn, O., Porciani, C., Carollo,
  C.~M., \& Dekel, A.\ 2007, \mnras, 375, 489

\bibitem[Hahn et al.(2007b)]{hahn2} Hahn, O., Carollo, C.~M.,
  Porciani, C., \& Dekel, A.\ 2007, \mnras, 381, 41

\bibitem[\protect\citeauthoryear{Hunt \& Malkan}{1999}]{hunt99} Hunt
  L.~K., Malkan M.~A., 1999, ApJ, 516, 660

\bibitem[\protect\citeauthoryear{Jogee, Scoville, \&
    Kenney}{2005}]{jogee05} Jogee S., Scoville N., Kenney J.~D.~P.,
  2005, ApJ, 630, 837

\bibitem[\protect\citeauthoryear{Knapen et al.}{1995}]{knapen95}
  Knapen J.~H., Beckman J.~E., Heller C.~H., Shlosman I., de Jong
  R.~S., 1995, ApJ, 454, 623

\bibitem[\protect\citeauthoryear{Knapen, Shlosman, \&
    Peletier}{2000}]{knapen00} Knapen J.~H., Shlosman I., Peletier
  R.~F., 2000, ApJ, 529, 93

\bibitem[\protect\citeauthoryear{Kormendy}{1982}]{kormendy82} Kormendy
  J., 1982, ApJ, 257, 75

\bibitem[\protect\citeauthoryear{Kormendy \&
    Kennicutt}{2004}]{kormendy04} Kormendy J., Kennicutt R.~C., 2004,
  ARA\&A, 42, 603

\bibitem[\protect\citeauthoryear{Laurikainen, Salo, \&
    Buta}{2004}]{laurikanen04} Laurikainen E., Salo H., Buta R., 2004,
  ApJ, 607, 103

\bibitem[\protect\citeauthoryear{Laurikainen et
    al.}{2007}]{laurikanen07} Laurikainen E., Salo H., Buta R., Knapen
  J.~H., 2007, MNRAS, 381, 401

\bibitem[\protect\citeauthoryear{Lee et al.}{2012}]{lee12} Lee G.-H.,
  Park C., Lee M.~G., Choi Y.-Y., 2012, ApJ, 745, 125

\bibitem[\protect\citeauthoryear{Li et al.}{2009}]{li09} Li C.,
  Gadotti D.~A., Mao S., Kauffmann G., 2009, MNRAS, 397, 726

\bibitem[Lintott et al.(2008)]{lintott1} Lintott, C.~J., Schawinski,
  K., Slosar, A., et al.\ 2008, \mnras, 389, 1179

\bibitem[Lintott et al.(2011)]{lintott2} Lintott, C., Schawinski, K.,
  Bamford, S., et al.\ 2011, \mnras, 410, 166

\bibitem[\protect\citeauthoryear{{\L}okas}{2018}]{lokas} {\L}okas E.~L., 2018, ApJ, 857, 6

\bibitem[\protect\citeauthoryear{Lynden-Bell}{1979}]{lyndenbell}
  Lynden-Bell D., 1979, MNRAS, 187, 101

\bibitem[\protect\citeauthoryear{Marinova \&
    Jogee}{2007}]{marinovajogee} Marinova I., Jogee S., 2007, ApJ,
  659, 1176

\bibitem[\protect\citeauthoryear{Marinova et al.}{2012}]{marinova12}
  Marinova I., Jogee S., Weinzirl T., Erwin P., Trentham N., Ferguson
  H.~C., Hammer D., et al., 2012, ApJ, 746, 136

\bibitem[\protect\citeauthoryear{Mart{\'\i}nez \&
    Muriel}{2011}]{martinez11} Mart{\'\i}nez H.~J., Muriel H., 2011,
  MNRAS, 418, L148

\bibitem[\protect\citeauthoryear{Martinez-Valpuesta et
    al.}{2017}]{valpuesta} Martinez-Valpuesta I., Aguerri J.~A.~L.,
  Gonz{\'a}lez-Garc{\'\i}a A.~C., Dalla Vecchia C., Stringer M., 2017,
  MNRAS, 464, 1502

\bibitem[\protect\citeauthoryear{M{\'e}ndez-Abreu,
    S{\'a}nchez-Janssen, \& Aguerri}{2010}]{mendez10} M{\'e}ndez-Abreu
  J., S{\'a}nchez-Janssen R., Aguerri J.~A.~L., 2010, ApJL, 711, L61

\bibitem[\protect\citeauthoryear{M{\'e}ndez-Abreu et
    al.}{2012}]{mendez12} M{\'e}ndez-Abreu J., S{\'a}nchez-Janssen R.,
  Aguerri J.~A.~L., Corsini E.~M., Zarattini S., 2012, ApJL, 761, L6

\bibitem[\protect\citeauthoryear{Pandey \& Sarkar}{2017}]{pandey17}
  Pandey B., Sarkar S., 2017, MNRAS, 467, L6

\bibitem[\protect\citeauthoryear{Planck Collaboration, et
    al.}{2018}]{planck18} Planck Collaboration, et al., 2018, arXiv,
  arXiv:1807.06209

\bibitem[\protect\citeauthoryear{Sarkar \& Pandey}{2020}]{sarkar20}
  Sarkar S., Pandey B., 2020, MNRAS, 497, 4077
  
\bibitem[\protect\citeauthoryear{Schwarz}{1981}]{schwarz81} Schwarz
  M.~P., 1981, ApJ, 247, 77
 
\bibitem[Shannon(1948)]{shannon48} Shannon, C. E. \ 1948, Bell System
  Technical Journal, 27, 379-423, 623-656

\bibitem[\protect\citeauthoryear{Sheth et al.}{2005}]{sheth05} Sheth
  K., Vogel S.~N., Regan M.~W., Thornley M.~D., Teuben P.~J., 2005,
  ApJ, 632, 217

\bibitem[\protect\citeauthoryear{Shlosman, Frank, \&
    Begelman}{1989}]{shlosman89} Shlosman I., Frank J., Begelman
  M.~C., 1989, Nature, 338, 45

\bibitem[\protect\citeauthoryear{Skibba et al.}{2012}]{skibba12}
  Skibba R.~A., Masters K.~L., Nichol R.~C., Zehavi I., Hoyle B.,
  Edmondson E.~M., Bamford S.~P., et al., 2012, MNRAS, 423, 1485

\bibitem[\protect\citeauthoryear{Thompson}{1981}]{thomson81} Thompson
  L.~A., 1981, ApJL, 244, L43

\bibitem[\protect\citeauthoryear{Toomre}{1964}]{toomre64} Toomre A.,
  1964, ApJ, 139, 1217

\bibitem[\protect\citeauthoryear{van den Bergh}{2002}]{vandenbergh}
  van den Bergh S., 2002, AJ, 124, 782

\bibitem[\protect\citeauthoryear{Wegg, Gerhard, \&
    Portail}{2015}]{wegg} Wegg C., Gerhard O., Portail M., 2015,
  MNRAS, 450, 4050

\bibitem[\protect\citeauthoryear{Weinberg}{1985}]{weinberg85} Weinberg
  M.~D., 1985, MNRAS, 213, 451

\bibitem[White \& Rees(1978)]{white78} White, S.~D.~M., \& Rees,
  M.~J.\ 1978, \mnras, 183, 341

\bibitem[\protect\citeauthoryear{Willett et al.}{2013}]{willet13}
  Willett K.~W., Lintott C.~J., Bamford S.~P., Masters K.~L., Simmons
  B.~D., Casteels K.~R.~V., Edmondson E.~M., et al., 2013, MNRAS, 435,
  2835

\bibitem[\protect\citeauthoryear{York, et al.}{2000}]{york00} York
  D.~G., et al., 2000, AJ, 120, 1579
  
  
\end{thebibliography}
\end{document}